\documentclass[12pt]{article}
\usepackage{amsfonts}
\usepackage{amsmath,amssymb,bm}

\begin{document}

\title{Vacuum static compactified wormholes in eight-dimensional Lovelock
theory}
\author{Fabrizio Canfora$^{1}$ \thanks{%
canfora AT cecs.cl}, Alex Giacomini$^{2,1}$ \thanks{%
giacomini AT cecs.cl}  \\
\\
{\small $^{1}$\textit{Centro de Estudios Cient\'{\i}ficos} (\textit{CECS),
Valdivia, Chile}}\\
{\small $^{2}$\textit{Instituto de Fisica, Facultad de Ciencias, Universidad
Austral de Chile, Valdivia, Chile}}\\
}
\maketitle

\begin{abstract}
In this paper new exact solutions in eight dimensional Lovelock theory will
be presented. These solutions are vacuum static wormhole, black hole and
generalized Bertotti-Robinson space-times with nontrivial torsion. All the
solutions have a cross product structure of the type $M_{5}\times \Sigma _{3}
$ where $M_{5}$ is a five dimensional manifold and $\Sigma _{3}$ a compact
constant curvature manifold. The wormhole is the first example of a smooth
vacuum static Lovelock wormhole which is neither Chern-Simons nor
Born-Infeld. It will be also discussed how the presence of torsion affects
the "navigableness" of the wormhole for scalar and spinning particles. It
will be shown that the wormhole with torsion may act as "geometrical
filter": a very large torsion may "increase the traversability" for scalars
while acting as a "polarizator" on spinning particles. This may have
interesting phenomenological consequences.


\end{abstract}

\section{Introduction}

Since the ideas of Kaluza and Klein and with the advent of string theory,
the possibility to have extra-dimensions became one of the most promising
possibility to extend the standard model of particles physics. Higher
dimensional theories of gravity may present some new features which are
absent in four dimensions. Indeed in four dimensions the only gravitational
action that can be built from curvature invariants leading to second order
equations in the metric is the Einstein-Hilbert action. The situation
changes in higher dimensions. In five dimensions one can add for example a
Gauss-Bonnet term to the action which is quadratic in the curvature and
leads to second order equations. In higher dimensions one can add higher
curvature powers to the action. Such higher curvature power theories leading
to second order equations for the metric are known as Lovelock theories \cite%
{Lovelock} (see, for pedagogical reviews on Lovelock gravities, \cite{Gaston}
\cite{TZ99} \cite{Za05} ).

Unlike General Relativity, the Lovelock equations of motion do not
imply the vanishing of torsion. However, the Lovelock equations of
motions put very strong constraints on the torsion so that it is
very difficult to find exact solutions with torsion. The only case
where such consistency conditions are automatically satisfied is
when in odd dimensions the coupling constants are tuned in such a
way that the theory becomes a Chern-Simons theory (see, for
instance, \cite{TZ99} \cite{Za05}). In this case the theory
possesses only one maximally symmetric vacuum as well as an enhanced
gauge symmetry. Because of the degeneracy of the theory it is easier
to find vacuum solutions with nontrivial torsion (however, often the
torsion turns out to be pure gauge). Following a suggestion in
\cite{Ca07} based on an analogy with BPS states in Yang-Mills
theory, the first vacuum solution with torsion in a non-Chern-Simons
theory was found in \cite{CGW07}. This solution is the purely
gravitational analogue of the Bertotti-Robinson space-time. The
crucial point of the construction was to make the ansatz of a
torsion concentrated on a three-dimensional sub-manifold (according
to the "BPS prescription" of \cite{Ca07}) so that many of the
torsion equations are identically satisfied. It is therefore
interesting to see if also more general solutions than the one
mentioned can be found in the non-Chern-Simons case as e.g. black
holes or wormholes. Indeed a black hole solution with the "BPS
torsion" in five dimension turns out to be Chern-Simons
\cite{CGTW07}.

An interesting possibility worth to be further analyzed is to try to
generalize the construction of \cite{Ca07}, \cite{CGW07} and
\cite{CGTW07} in higher dimensions. The even-dimensional case is
particularly interesting since, in these cases, one avoids in part
the degeneracies of Chern-Simons theory as there is no enhanced
gauge symmetry in this case. Thus, we will consider the
eight-dimensional case: one can add another 3-D compact
sub-manifolds which implies that there are now two possibilities to
put "BPS torsion". We will search for solutions in eight dimensions
with the structure of $M_{5}\times \Sigma _{3}$ where $M_{5}\equiv
M_{2}\times F(r)N_{3}$ is a five dimensional manifold, where $M_{2}$
plays the role of an $r-t$ plane, $N_{3}$ is a constant curvature
manifold (which we will call base manifold) and $F(r)$ is a warp
factor. $\Sigma _{3}$ is another compact constant curvature manifold
that plays the role of a compactified 3-D space. Of course, such
eight dimensional solutions are non-Chern-Simons as in even
dimensions Chern-Simons theories cannot exist. In even dimensions
Born-Infeld gravity is the most similar theory to Chern-Simons
gravity since this theory admits only one maximally symmetric vacuum
(see, for instance, \cite{TZ99} \cite{Za05}). It will therefore be
of special interest to find exact solutions with nontrivial torsion
in the non-Born-Infeld case.

Lovelock theories admit static vacuum wormhole solutions (static
wormholes in gravity in dimension higher than four were found in
\cite {Tulin}, \cite {Tulin3} in the case of Chern-Simons theory.
Indeed one can see that for the Einstein-Gauss-Bonnet theory vacuum
wormholes exist in any dimension provided that there is a unique
maximally symmetric vacuum \cite{tulincito}). Wormhole solutions are
interesting in that they heavily try out the geometric structure of
the theory. Indeed in four dimensions, static wormhole solutions
cannot exist in vacuum and the matter that sustains such solutions
violates the energy conditions (for a nice review see
\cite{VisserL}; in Einstein-Gauss-Bonnet theory there exist
non-vacuum wormhole solutions whose energy-momentum tensors respect
the energy conditions \cite{maeda}). It is therefore interesting to
search for smooth static vacuum wormhole solutions in more general
Lovelock theories which have a quite rich dynamical content. We will
construct in this paper eight dimensional exact static vacuum
solutions which have the structure of $M_{5}\times \Sigma _{3}$.
Some of these exact static vacuum solutions can be considered as
effective five dimensional wormholes with three compactified extra
dimensions playing a "spectator" role. We will show that these
solutions can carry non-trivial torsion. A remarkable feature of the
present construction is that the presence of torsion affects the
"navigableness" of the wormhole: torsion has quite different effects
on scalars and spinning particles. A large torsion improves the
"navigableness" for scalars while acting on spinning particles, in a
sense, as a "polarizator". In very much the same way as the coupling
of a magnetic field with a Fermion, the coupling of torsion with
angular momenta favours angular momenta "pointing in the same
direction as" torsion: a throat with torsion can act as a filter for
particles which have not the right polarization (with respect to the
background torsion).

The same geometric structure of $M_{5}\times \Sigma _{3}$ can also
support effective five-dimensional black hole solutions which can
have non-trivial torsion. Another simple class of solutions with
torsion has the structure of $M_{2}\times S_{3}\times S_{3}$ where
$M_{2}$ is a two-dimensional constant curvature Lorentzian manifold
and can be seen as a generalized Bertotti-Robinson space-time.

The structure of the paper will be the following: In the second
section we will give a short review of eight dimensional Lovelock
theory. In third and fourth section the curvature and the torsion
and the corresponding equations of motion for a manifold of the form
$M_{5}\times \Sigma _{3}$ will be discussed. In the fifth and sixth
sections the wormhole solutions and their "navigableness" will be
analyzed. Then the black hole solutions and the generalized
Bertotti-Robinson solutions will be shortly described. Eventually,
some conclusive remarks will be given.

\section{Eight dimensional Lovelock theory}

The most general Lovelock action in 8-D in reads
\begin{gather}
I=\int \epsilon _{ABCDEFGH}\left( \frac{c_{0}}{8}%
e^{A}e^{B}e^{C}e^{D}e^{E}e^{F}e^{G}e^{H}+\frac{c_{1}}{6}%
R^{AB}e^{C}e^{D}e^{E}e^{F}e^{G}e^{H}\right.  \notag \\
\left. +\frac{c_{2}}{4}R^{AB}R^{CD}e^{E}e^{F}e^{G}e^{H}+\frac{c_{3}}{2}%
R^{AB}R^{CD}R^{EF}e^{G}e^{H}\right)  \label{lovelock8}
\end{gather}%
where in the second order formalism the quadratic (Gauss-Bonnet)
term proportional to $c_2$ reads
\begin{equation}
R^{AB}R^{CD}e^{E}e^{F}e^{G}e^{H}\epsilon _{ABCDEFGH}= R^2 - 4R^{\mu
\nu}R_{\mu \nu} + R^{\alpha \beta \mu \nu}R_{\alpha \beta \mu \nu}
\end{equation}
and the cubic term  proportional to $c_3$ reads
\begin{gather}
R^{AB}R^{CD}R^{EF}e^{G}e^{H}\epsilon _{ABCDEFGH}=R^3+ 3RR^{\mu \nu
\alpha \beta}R_{\alpha \beta \mu \nu} -12RR^{\mu \nu}R_{\mu
\nu}\notag \\
+24 R^{\mu \nu \alpha \beta}R_{\alpha \mu}R_{\beta \nu} + 16R^{\mu
\nu}R_{\nu \alpha}R^{\alpha}_{\mu}+ 24R^{\mu \nu \alpha
\beta}R_{\alpha \beta \nu \rho}R^{\rho}_{\mu} \notag \\
 + 8R^{\mu \nu}_{\phantom{aa} \alpha \rho}R^{\alpha \beta}_{\phantom{aa} \nu \sigma}R^{\rho \sigma}_{\phantom{aa} \mu
 \beta}+ 2R_{\alpha \beta \rho \sigma}R^{\mu \nu \alpha \beta}R^{\rho \sigma}_{\phantom{aa} \mu \nu}
\end{gather}
where $R^{\alpha \beta \mu \nu}$, $R^{\mu \nu}$ and $R$ are
respectively the Riemann tensor the Ricci tensor and the Ricci
scalar in the second order formalism.

The equations of motion variating the action with respect to the vielbein
are
\begin{gather}
\epsilon _{H}=\epsilon _{ABCDEFGH}\left(
c_{0}e^{A}e^{B}e^{C}e^{D}e^{E}e^{F}e^{G}+c_{1}R^{AB}e^{C}e^{D}e^{E}e^{F}e^{G}\right.
\notag \\
\left. +c_{2}R^{AB}R^{CD}e^{E}e^{F}e^{G}+c_{3}R^{AB}R^{CD}R^{EF}e^{G}\right)
=0  \label{equ}
\end{gather}%
And varying with respect to the spin connection one obtains%
\begin{equation*}
\epsilon _{GH}=\epsilon _{ABCDEFGH}T^{A}\left(
c_{1}e^{B}e^{C}e^{D}e^{E}e^{F}+\right.
\end{equation*}
\begin{equation}
\left. +2c_{2}R^{BC}e^{D}e^{E}e^{F}+3c_{3}R^{BC}R^{DE}e^{F}\right) =0.
\label{equt}
\end{equation}%
where the torsion two form $T^A=de^A + \omega^{A}_{\phantom{a}B}
e^B$ is related to the antisymmetric part of the Christoffel symbols
by
\begin{equation}
T^{\lambda}_{\phantom{a}\mu \nu} \equiv e_A^{\phantom{a}\lambda}T^A
_{\phantom{a}\mu \nu}= 2 \Gamma^{\lambda}_{\phantom{a}[\mu \nu]}
\end{equation}
In even dimension a Lovelock theory is called Born-Infeld theory if
the coupling constants are tuned in such a way that the theory
admits only one maximally symmetric vacuum. In eight dimensions the
Born-Infeld Lagrangian
has the form%
\begin{equation*}
I_{BI}=\int \epsilon _{ABCDEFGH}(R^{AB}+\Lambda e^{A}e^{B})\cdot
\end{equation*}
\begin{equation*}
\cdot (R^{CD}+\Lambda e^{C}e^{D})(R^{ef}+\Lambda e^{E}e^{F})(R^{GH}+\Lambda
e^{G}e^{H}).
\end{equation*}%
If we compare this action with (\ref{lovelock8}) it is easy to find the
tuning of the coupling constants leading to Born-Infeld action
\begin{equation}
c_{2}^{2}=3c_{1}c_{3};\ \ \;\;\;c_{1}^{2}=3c_{0}c_{2}.  \label{BItuning}
\end{equation}

The Born-Infeld Lagrangians in even dimensions and the Chern-Simons
Lagrangians in odds dimensions share some interesting features (see,
for instance, \cite{TZ99}, \cite{Za05}). First of all, in both cases
there is a unique maximally symmetric vacuum. One can also argue
that both types of Lagrangians have some degree of degeneracy in the
sense that it may happen that the equations of motion leave
undetermined some of the metric functions. Thus, one may expect that
to construct static vacuum wormholes in the Born-Infeld and
Chern-Simons cases can be quite easier than in the generic case (up
to now, the only static vacuum wormhole is the one in \cite {Tulin},
\cite{Tulin3}, \cite{tulincito} in which a certain degree of
degeneracy is manifest). In the generic case the wormhole structure
imposes very strong constraints on the metric
functions and on the energy momentum tensor (see, for instance, \cite%
{VisserL}). In five dimensions, only recently it appeared the first
stationary Ricci flat wormhole \cite{Lu}. For these reasons, we will
construct here static vacuum wormhole spacetimes in non-Born-infeld cases
(namely, when the conditions in Eq. (\ref{BItuning}) are not fulfilled). We
will also consider the effects of torsion which, in higher dimensions, is
generically different from zero. In the generic non-Chern-Simons Lovelock
case, the equations of motions for the torsion are very restrictive and
until very recently, no exact solution with torsion was known. The first was
discovered in \cite{CGW07} using the ansatz for the torsion (inspired by an
analogy with gauge theory first proposed in \cite{Ca07})%
\begin{equation}
T^{i}=K(r)\epsilon ^{ijk}e_{j}e_{k}.  \label{marameo}
\end{equation}

\section{Ansatz of $M_5 \times S_3$ manifold}

We make the ansatz for the metric
\begin{equation*}
ds^{2}=-\left( f\left( r\right) \right) ^{2}dt^{2}+\frac{dr^{2}}{\left(
g\left( r\right) \right) ^{2}}+r^{2}d\Sigma _{1}^{2}+d\Sigma _{2}^{2}
\end{equation*}%
where $\Sigma _{1}$ (which, from now on, we will call "\textit{base manifold}%
") and $\Sigma _{2}$ are two constant curvature 3-D manifolds. We can choose
the vielbein as (using the indices $i$, $j$, $k$ for $\Sigma _{1}$ and $a$, $%
b$, $c$ for $\Sigma _{2}$)
\begin{equation}
\begin{array}{cccc}
e^{0}=fdt, & e^{1}=\frac{dr}{g}, & e^{i}=r\hat{e}^{i}, & e^{a}=\hat{e}^{a}%
\end{array}
\label{vielbein}
\end{equation}%
so that the Riemannian part of the connection is%
\begin{equation}
\begin{array}{cccc}
\omega _{\ 1}^{0}=g\frac{\left( f\right) ^{\prime }}{f}e^{0}, & \omega _{\ \
1}^{i}=\frac{g}{r}e^{i}, & \omega _{\ \ j}^{i}=\widehat{\omega }_{\ \ j}^{i},
& \omega _{\ \ b}^{a}=\widehat{\omega }_{\ \ b}^{a},%
\end{array}
\label{conn}
\end{equation}%
where we used the notation
\begin{equation*}
f^{\prime }=\partial _{r}f,\ \ \ \ g^{\prime }=\partial _{r}g,
\end{equation*}%
and
\begin{equation*}
\omega _{\ \ 1}^{a}=0=\omega _{\ \ 0}^{a}=\omega _{\ \ 0}^{i}=0=\omega _{\ \
a}^{i}
\end{equation*}%
where $\hat{e}^{i}$ and $\hat{e}^{a}$ are the intrinsic vielbeins
respectively of the base manifold $\Sigma _{1}$ and of $\Sigma _{2}$ and $%
\widehat{\omega }_{\ \ j}^{i}$ and $\widehat{\omega }_{\ \ b}^{a}$
are the intrinsic spin connections of the base manifold $\Sigma
_{1}$ and of $\Sigma _{2}$. For the torsion we make the following
ansatz
\begin{equation}
\begin{array}{ccc}
T^{0}=T^{1}=0, & T^{i}=K_{1}(r)\epsilon ^{ijk}e_{j}e_{k}, &
T^{a}=K_{2}(r)\epsilon ^{abc}e_{b}e_{c}%
\end{array}
\label{anstor}
\end{equation}%
so that the contorsion is
\begin{equation}
\begin{array}{cc}
K^{ij}=-K_{1}(r)\epsilon ^{ijk}e_{k},\ \ \  & K^{ab}=-K_{2}(r)\epsilon
^{abc}e_{c},%
\end{array}
\label{cont}
\end{equation}%
and the total connection is%
\begin{equation}
\left( \omega _{tot}\right) ^{AB}=\omega ^{AB}+K^{AB}.  \label{toconn}
\end{equation}%
With this ansatz the total curvature two forms (which includes torsion
effects) take the form
\begin{equation*}
R^{0a}=R^{1a}=R^{ia}=0
\end{equation*}%
\begin{equation*}
R^{01}=-\frac{g}{f}(gf^{\prime })^{\prime }e^{0}e^{1}
\end{equation*}%
\begin{equation*}
R^{0i}=-\frac{g^{2}f^{\prime }}{fr}e^{0}e^{i}
\end{equation*}%
\begin{equation*}
R^{1i}=-\frac{gg^{\prime }}{r}e^{0}e^{i}-\frac{1}{r}T^{i}
\end{equation*}%
\begin{equation*}
R^{ij}=\left( \frac{\gamma -g^{2}}{r^{2}}\right) e^{i}e^{j}-\frac{d(rK_{1})}{%
r}\epsilon ^{ijk}e_{k}-K_{1}^{2}e^{i}e^{j}
\end{equation*}%
\begin{equation*}
R^{ab}=\eta e^{a}e^{b}-d(K_{2})\epsilon ^{abc}e_{c}-K_{2}^{2}e^{a}e^{b}
\end{equation*}%
where $\gamma $ and $\eta $ are the intrinsic scalar curvatures of $\Sigma
_{1}$ and $\Sigma _{2}$ respectively.

It is worth to stress here that torsion can be divided into its irreducible
components: it is trivial to see that when torsion has the form in Eq. (\ref%
{anstor}) it is fully skew-symmetric so that both its trace and its
symmetric parts vanish: in these cases (see for instance \cite{Shapiro}) one
says that only the axial part of the torsion is present.

\section{Equations of motion}

It is easy to see that in the equations of motions Eq. (\ref{equ}), in all
the terms in which it appears $R^{1i}$ the non-Riemannian part proportional
to $T^{i}$ do not contribute to the equations of motion due to the
identities satisfied by ansatz used for the torsion (see \cite{Ca07}). In
order to satisfy the equations of motion one must have
\begin{eqnarray}
d(rK_{1}) &=&0;\;\;\;dK_{2}=0\Rightarrow  \label{tor0} \\
T^{i} &=&\frac{\delta _{(1)}}{r}\epsilon ^{ijk}e_{j}e_{k},  \label{tor1} \\
T^{a} &=&K_{2}\epsilon ^{abc}e_{b}e_{c}  \label{tor2}
\end{eqnarray}%
where $\delta _{(1)}$ and $K_{2}$ are constants.

It is worth to stress here the following point: the constants $\gamma $ and $%
\eta $\ can be rescaled to $-1$, $0$ and $1$ (according to the
scalar curvatures of $\Sigma _{1}$ and $\Sigma _{2}$). However, the
"torsion" constants $\delta _{(1)}$ and $K_{2}$ cannot be rescaled
away, they can take any real values (compatible with the equations
of motion) since they are
true integration constants representing the strength of the torsion in the $%
i $ and the $a$ directions. Thus, as in \cite{CGW07}, the presence of
torsion will be manifest directly in the metric: this is unlike the
five-dimensional Chern-Simons case in which in the half BPS black hole
constructed in \cite{CGTW07}, torsion manifests itself mainly in the Killing
spinor equation.

It is convenient to introduce the following definitions in order to simplify
the notation
\begin{equation}
R^{01}\equiv Fe^{0}e^{1};\;\;\;R^{0i}\equiv Ae^{0}e^{i};\;\;\;R^{1i}\equiv
Be^{1}e^{i}-\frac{1}{r}T^{i}  \label{def1}
\end{equation}%
\begin{equation}
R^{ij}\equiv D(r)e^{i}e^{j};\;\;\;R^{ab}\equiv \tilde{\eta}%
e^{a}e^{b}\Rightarrow  \label{def2}
\end{equation}%
\begin{eqnarray}
A &=&-\frac{g^{2}f^{\prime }}{fr},\ \ \ \ B=-\frac{gg^{\prime }}{r},\ \ \
\label{def1.25} \\
F &=&-\frac{g}{f}(gf^{\prime })^{\prime },\ \ \ \ D=\frac{\tilde{\gamma}%
-g^{2}}{r^{2}},  \label{def1.5}
\end{eqnarray}%
where $\tilde{\gamma}=\gamma -(\delta _{(1)})^{2}$ and $\tilde{\eta}=\eta
-K_{2}^{2}$ are the effective curvatures (shifted by the fluxes of torsion).
We can now write the equations of motion for our ansatz:%
\begin{equation*}
\epsilon _{0}=B\left[ 20c_{1}+4c_{2}\left( D+3\tilde{\eta}\right) +12c_{3}%
\tilde{\eta}D\right]
\end{equation*}%
\begin{equation}
+\left[ \ 140c_{0}+20c_{1}D+20c_{1}\tilde{\eta}+12c_{2}\tilde{\eta}D\right]
=0,  \label{equ2}
\end{equation}%
\begin{equation*}
\epsilon _{1}=A\left[ 20c_{1}+4c_{2}\left( D+3\tilde{\eta}\right) +12c_{3}%
\tilde{\eta}D\right]
\end{equation*}%
\begin{equation}
+\left[ \ 140c_{0}+20c_{1}D+20c_{1}\tilde{\eta}+12c_{2}\tilde{\eta}D\right]
=0,  \label{equ3}
\end{equation}%
\begin{gather}
\epsilon _{i}=F\left( 20c_{1}+4c_{2}D+12c_{2}\tilde{\eta}+12c_{3}\tilde{\eta}%
D\right) +A\left( 40c_{1}+24c_{2}\tilde{\eta}\right)  \notag \\
+AB\left( 8c_{2}+24c_{3}\tilde{\eta}\right) +B\left( 40c_{1}+24c_{2}\tilde{%
\eta}\right)  \notag \\
+\left( 420c_{0}+20c_{1}D+60c_{1}\tilde{\eta}+12c_{2}\tilde{\eta}D\right) =0,
\label{equ4}
\end{gather}%
\begin{equation*}
\epsilon _{a}=F\left( 20c_{1}+12c_{2}D+4c_{2}\tilde{\eta}+12c_{3}\tilde{\eta}%
D\right)
\end{equation*}

\begin{gather}
+A\left( 60c_{1}+12c_{2}\tilde{\eta}+12c_{2}D+12c_{3}\tilde{\eta}D\right)
\notag \\
+AB\left( 24c_{2}+24c_{3}\tilde{\eta}\right) +B\left( 60c_{1}+12c_{2}\tilde{%
\eta}+12c_{2}D+12c_{3}\tilde{\eta}D\right)  \notag \\
+\left( 420c_{0}+60c_{1}D+20c_{1}\tilde{\eta}+12c_{2}\tilde{\eta}D\right) =0.
\label{equ5}
\end{gather}%
As far as the equations in which torsion appears explicitly is concerned, a
very nice feature of the ansatz for the torsion (\ref{marameo}) is that
almost all the equations (\ref{equt}) are identically fulfilled. The only
components of the torsion equations (\ref{equt}) which do not vanish
identically because of the ansatz of the torsion read
\begin{equation}
\epsilon _{ij}=0\Rightarrow \delta _{\left( 1\right) }\left( F\left(
4c_{2}+12c_{3}\tilde{\eta}\right) +20c_{1}+12c_{2}\tilde{\eta}\right) =0,
\label{equ6}
\end{equation}%
\begin{gather}
\epsilon _{ab}=0\Rightarrow K_{2}\left( F\left( 4c_{2}+12c_{3}D\right)
+B\left( 12c_{2}+12c_{3}D\right) +A\left( 12c_{2}+12c_{3}D\right) \right.
\notag \\
\left. +24c_{3}AB+20c_{1}+12c_{2}D\right) =0  \label{equ7}
\end{gather}%
It is important to notice that Eqs. (\ref{equ2}) and (\ref{equ3}) are equal
provided $B$ is exchanged with $A$. Therefore in order to be compatible
there are two possibilities:

the first is $A=B$ which implies $f=g$ and so it corresponds to black hole
like solutions.

The second possibility appears when in the two square brackets in Eqs. (\ref%
{equ2}) and (\ref{equ3}) are zero separately leaving open the possibility to
have wormhole solutions.

We will discuss the two possibilities separately.

\section{Wormholes}

In order to have $f\neq g$ the two square brackets in Eqs. (\ref{equ2}) and (%
\ref{equ3}) must be zero separately: the reason is that in the generic case
the consistency of Eqs. (\ref{equ2}) and (\ref{equ3}) imply $A=B$ (where $A$
and $B$ are defined in Eqs. (\ref{def1}), (\ref{def1.25}) and (\ref{def1.5}%
)) and then $f=g$. Thus, in order to avoid this "no-go argument" for the
appearance of wormhole, one has to ask that%
\begin{eqnarray}
\left( 4c_{2}+12c_{3}\tilde{\eta}\right) D(r) &=&-\left( 20c_{1}+12c_{2}%
\tilde{\eta}\right) ,  \label{def1.01} \\
\left( 20c_{1}+12c_{2}\tilde{\eta}\right) D(r) &=&-\left( 140c_{0}+20c_{1}%
\tilde{\eta}\right) .  \label{def1.02}
\end{eqnarray}%
The consistency condition of Eqs. (\ref{def1.01}) and (\ref{def1.02}) is
\begin{equation}
\frac{20c_{1}+12c_{2}\tilde{\eta}}{4c_{2}+12c_{3}\tilde{\eta}}=\frac{%
140c_{0}+20c_{1}\tilde{\eta}}{20c_{1}+12c_{2}\tilde{\eta}}  \label{def2.5}
\end{equation}%
which is nothing but the requirement that Eqs. (\ref{equ2}) and (\ref{equ3})
should be identically satisfied no matter the values of $A$ and $B$. This
condition fixes the function $D(r)$ to be a constant $D_{0}$ as follows
\begin{equation}
D(r)\equiv D_{0}=-\frac{20c_{1}+12c_{2}\tilde{\eta}}{4c_{2}+12c_{3}\tilde{%
\eta}}=-\frac{140c_{0}+20c_{1}\tilde{\eta}}{20c_{1}+12c_{2}\tilde{\eta}}
\label{def3}
\end{equation}%
Using the definition of $D(r)$ in Eqs. (\ref{def2}) and (\ref{def1.5}), we
obtain the form of $g^{2}$
\begin{equation}
g^{2}=-D_{0}r^{2}+\tilde{\gamma}  \label{def4}
\end{equation}%
In the case of vanishing torsion (in which $\tilde{\gamma}=\gamma $ and $%
\tilde{\eta}=\eta $), in order for the metric to represent a static
wormhole, a necessary condition is that the $g_{rr}$ metric component has
the form in Eq. (\ref{def4}) with both $D_{0}$ and $\gamma $ negative (see,
for instance, \cite{VisserL},\ \cite{Tulin}).

There are two possibilities:

the first possibility is to satisfy the constraint (\ref{def2.5}) so that
the $g_{rr}$ component is determined by Eq. (\ref{def4}).

The second possibility is when the round brackets in Eqs. (\ref{def1.01})
and (\ref{def1.02}) vanish identically leaving $D$ indeterminate. This
happens for the following tuning of the couplings and $\tilde{\eta}$
\begin{equation}
5c_{1}+3c_{2}\tilde{\eta}=0;\;\ \ \;\;c_{2}+3c_{3}\tilde{\eta}=0;\;\;\ \
\;7c_{0}+c_{1}\tilde{\eta}=0  \label{degeneracy1}
\end{equation}%
which imply that%
\begin{equation*}
\tilde{\eta}=-\frac{c_{2}}{3c_{3}}
\end{equation*}%
together with two relations involving only the couplings
\begin{equation}
\;c_{2}^{2}=5c_{1}c_{3};\ \ \;\;\;5c_{1}^{2}=21c_{2}c_{0}
\label{degeneracy2}
\end{equation}%
At a first glance, one would expect that such a strong degeneracy condition
should correspond to the Born-Infeld case which is the even-dimensional
analogous of the Chern-Simons Lagrangians (see, for instance, \cite{TZ99}
\cite{Za05}). Often, this implies that the field equations manifest a huge
degeneracy in such a way that the metric (representing a given exact
solution) may have arbitrary functions left completely undetermined by the
field equations themselves. As a matter of fact, the first exact vacuum
solution representing a static wormhole has been found in the Chern-Simons
case \cite{Tulin}. Thus, when searching vacuum static wormholes, one may
think that the only hope to find them is in the cases of Born-Infeld or
Chern-Simons. Remarkably enough, the relations above \textit{do not
correspond to the Born-Infeld tunings} (\ref{BItuning}). We will treat the
degenerate and non-degenerate case separately

\subsection{Degenerate wormhole}

In this and in the following subsections we will consider wormholes
characterized by the fact that
\begin{equation*}
f=const,\ \ \ \ \ D(r)=const\equiv D_{0}.
\end{equation*}%
where $D(r)$ is defined in Eq. (\ref{def2}). The reason is that these are
the simplest and most elegant wormholes in which one can see in the clearest
possible way the different effects of torsion on the "throat
"navigableness"" of scalar and spinning particles. As it will be explained
in the next sub-section, in the generic case $f$ satisfies a
hypergeometric-like equation. Thus, wormhole solutions correspond to
hypergeometric functions without zeros in $r$. However, from the qualitative
point of view, wormhole solutions with a non-constant $f$ do not have new
features if compared to wormhole solutions with $f$ constant.

The easiest way to proceed is to first solve the torsion equations: the $ij$
component of the torsion equations (namely, Eq. (\ref{equ6})) is identically
satisfied because of the degeneracy condition in Eq. (\ref{degeneracy2}).
The $ab$ component of the torsion equations (that is, Eq. (\ref{equ7})) is
\begin{equation}
\epsilon _{ab}=12c_{3}D_{0}^{2}+24c_{2}D_{0}+20c_{1}=0,  \label{dewortor1}
\end{equation}%
This fixes the constant $D_{0}$ in terms of the coupling constants $c_{i}$.
When the degeneracy conditions in Eq. (\ref{degeneracy2}) hold, Eqs. (\ref%
{equ2}) and (\ref{equ3}) are identically satisfied; it also is immediate to
see that, due to the degeneracy conditions in Eq. (\ref{degeneracy2}), Eq. (%
\ref{equ4}) is identically satisfied as well. Only Eq. (\ref{equ5}) is left:
\begin{equation}
D_{0}^{2}(12c_{2}+12c_{3}\tilde{\eta})+D_{0}(120c_{1}+24c_{2}\tilde{\eta}%
)+(420c_{0}+20c_{1}\tilde{\eta})=0  \label{dewortor2}
\end{equation}%
Inserting the degeneracy conditions this becomes
\begin{equation}
8c_{2}D_{0}^{2}+80c_{1}D_{0}+280c_{0}=0  \label{dewortor3}
\end{equation}%
It is easy to check (once the conditions in Eq. (\ref{degeneracy2}) are
taken into account) that this quadratic equation in $D_{0}$ has the same
roots as Eq. (\ref{dewortor1}) and so they are compatible.

It is worth to notice that, unlike the static vacuum Chern-Simons worm-hole
of \cite{Tulin} in which there is a certain degree of degeneracy in the
metric\footnote{%
One can also see that even adding a "BPS" torsion of the form in Eq. (\ref%
{anstor}) to the static vacuum Chern-Simons worm-hole the degeneracy, in
general, is not removed.}, in the present degenerate case when there is a
non-vanishing torsion on the $\Sigma _{2}$ sub-manifold the indeterminacy is
completely lifted. However, in the zero torsion sector of the theory, when
conditions in Eq. (\ref{degeneracy2}) hold, the equations of motion would be
under-determined (since Eqs. (\ref{equ2}), (\ref{equ3}) and (\ref{equ4}) are
identically fulfilled). This makes manifest the important role of torsion in
removing degeneracies.

In order to display in a clear way the structure of the wormhole, the
following change of coordinates is useful (it is worth to remember here that
both $D_{0}$ and $\gamma -(\delta _{(1)})^{2}$ are negative in the case of a
wormhole):%
\begin{eqnarray}
\frac{dr^{2}}{g^{2}} &=&\frac{dr^{2}}{\left( -D_{0}\right) r^{2}-(\delta
_{(1)})^{2}+\gamma }=d\rho ^{2}\Rightarrow  \label{cam1} \\
r &=&r_{G}\cosh \left( \left( -D_{0}\right) ^{1/2}\rho \right)  \label{cam2}
\end{eqnarray}%
where $r_{G}$ will be defined in a moment (see Eq. (\ref{raga}) below).
Thus, the wormhole metric is%
\begin{eqnarray}
ds^{2} &=&-dt^{2}+d\rho ^{2}+r_{G}^{2}\cosh ^{2}\left( \left( -D_{0}\right)
^{1/2}\rho \right) d\Sigma _{1}^{2}+d\Sigma _{2}^{2},  \label{dewo1} \\
T^{i} &=&\frac{\delta _{(1)}}{r_{G}\cosh \left( \left( -D_{0}\right)
^{1/2}\rho \right) }\epsilon ^{ijk}e_{j}e_{k},\ \ \ T^{a}=K_{2}\epsilon
^{abc}e_{b}e_{c},  \label{dewo2}
\end{eqnarray}%
where the range of the coordinate $\rho $ extends from $-\infty $ to $%
+\infty $, $D_{0}$ is one of the roots of Eq. (\ref{dewortor3}), the throat
is located at $\rho =0$ and the throat radius $r_{G}$ is:
\begin{equation}
r_{G}=\left( \frac{(\delta _{(1)})^{2}-\gamma }{\left( -D_{0}\right) }%
\right) ^{1/2}.  \label{raga}
\end{equation}%
A wormhole metric  with flat $\rho -t $ plane, as the above one, in
four dimensions is a vacuum solution of conformal gravity
\cite{tempo}. The constant of integration $\delta _{(1)}$ (which
characterizes the strength of the torsion in the $i$ directions) is
not fixed by the equations of motion so that, by varying $\delta
_{(1)}$, one can obtain effective five dimensional wormholes of any
radius. It is worth to point out that in order
to have a wormhole solution the effective curvature of the base manifold $%
\Sigma _{1}$ which is given by $\tilde{\gamma}$ must be negative. Of course,
in the metric of the base manifold $d\Sigma _{1}^{2}$ (which is made out of
the vielbeins which do not receive torsion corrections in this framework) it
only enters the Riemannian curvature $\gamma $. This means that one can have
a base manifold with positive constant Riemannian curvature $\gamma $
provided a non-zero torsion concentrated on the base manifold makes $\tilde{%
\gamma}$ negative.

It is worth to recall here a known but important point (see
\cite{SUSY}, \cite{MM}, \cite{Tulin2}): in the case in which the
Riemannian curvature of the base manifold $\gamma $ is negative, in
order to get a wormhole instead
of $\Sigma _{1}$ itself, one has to consider the quotient $\widehat{\Sigma }%
_{1}$ of the base manifold by a freely acting discrete subgroup
$\Gamma $
(otherwise there would be only one asymptotic region)%
\begin{equation*}
\widehat{\Sigma }_{1}=\Sigma _{1}/\Gamma ;
\end{equation*}%
indeed, the local expression of the gravitational field is the same as with $%
\Sigma _{1}$.

In the case in which the base manifold has negative curvature the effective
5-D metric has locally (and hence also asymptotically on both side of the
throat for $\rho \rightarrow \pm \infty $) the form $R\times H_{4}$. In the
case in which the base manifold has positive constant curvature the
effective 5-D metric is only asymptotically locally of the form $R\times
H_{4}$. Thus, in both cases the asymptotic metric is the same on both side
of the throat. The only qualitative difference in the wormhole solutions in
which the metric function $f$ is not constant is that the asymptotic is, in
general, different on the two sides $\rho \rightarrow \pm \infty $.

\subsection{Non-degenerate wormholes}

In this non-degenerate case Eqs. (\ref{def2.5}), (\ref{def3}) and (\ref{def4}%
) imply that
\begin{equation*}
g^{2}=-D_{0}r^{2}+\gamma .
\end{equation*}%
It is trivial to see that $B=D_{0}$ ($B$ is defined in Eq. (\ref{def1})) and
that Eq. (\ref{equ4}) is identically satisfied due to Eq. (\ref{def2.5}).
The equation Eq. (\ref{equ5}) is a hyper-geometric like equation for $f$ (as
it can be checked by substituting the above expression of $g^{2}$ into Eqs. (%
\ref{def1}), (\ref{def2}) and (\ref{equ5})): thus, worm-hole solutions
correspond to hyper-geometric functions without zeros. The simplest solution
(which, nevertheless, manifests all the expected non-trivial features) is $%
f=const$. In this case $F=A=0$ and Eq. (\ref{equ5}) reduces to
\begin{equation}
D_{0}^{2}(12c_{2}+12c_{3}\tilde{\eta})+D_{0}(120c_{1}+24c_{2}\tilde{\eta}%
)+420c_{0}+20c_{1}\tilde{\eta}=0  \label{pol1}
\end{equation}%
Thus, in the zero torsion sector, one gets a wormhole provided the base
manifold $\Sigma _{1}$ is compact and of constant negative curvature (since $%
\gamma $ has to be negative). As it has been already explained, this can be
achieved with the procedure outlined in \cite{SUSY}, \cite{MM}, \cite{Tulin2}%
. Therefore, the vacuum static wormhole in the non-degenerate case is%
\begin{equation}
ds^{2}=-dt^{2}+d\rho ^{2}+\left( \frac{-\gamma }{\left( -D_{0}\right) }%
\right) \cosh ^{2}\left( \left( -D_{0}\right) ^{1/2}\rho \right) d\Sigma
_{1}^{2}+d\Sigma _{2}^{2},  \label{nondewo0}
\end{equation}%
where we have again used the transformation in Eq. (\ref{cam1}).

Also in this case in which the base manifold has negative curvature the
effective 5-D metric has locally (and hence also asymptotically on both side
of the throat for $\rho \rightarrow \pm \infty $) the form $R\times H_{4}$.
Thus, the asymptotic metric is the same on both side of the throat. Even in
this case, the wormhole solutions in which the metric function $f$ is not
constant only differ from the ones in Eq. (\ref{nondewo0}) in that the
asymptotic is, in general, different on the two sides of the throat $\rho
\rightarrow \pm \infty $.

The $ij$ component of the torsion equations would imply the condition $%
5c_{1}+3c_{2}\tilde{\eta}=0$ which is one of the degeneracy conditions in
Eq. (\ref{degeneracy2}) and therefore such a case is excluded here so that,
in this non-degenerate case, $T^{i}=0$. The $ab$ component (that is, Eq. (%
\ref{equ7})) gives
\begin{equation}
\epsilon _{ab}=12c_{3}D_{0}^{2}+24c_{2}D_{0}+20c_{1}=0  \label{pol2}
\end{equation}%
Thus, the wormhole metric reads%
\begin{eqnarray}
ds^{2} &=&-dt^{2}+d\rho ^{2}+\left( \frac{-\gamma }{\left( -D_{0}\right) }%
\right) \cosh ^{2}\left( \left( -D_{0}\right) ^{1/2}\rho \right) d\Sigma
_{1}^{2}+d\Sigma _{2}^{2},  \label{nondewo1} \\
T^{a} &=&K_{2}\epsilon ^{abc}e_{b}e_{c},  \label{nondewo2}
\end{eqnarray}%
where it is worth to stress that, unlike the previous degenerate case, the
throat\ radius is fixed by the coupling constant of the theory through $%
D_{0} $ (see Eq. (\ref{d02}) below).

It is important to notice that the Eqs. (\ref{def2.5}), (\ref{def3}), (\ref%
{pol1}) and (\ref{pol2}) taken together imply an extremely awkward set of
constraints on the coupling constants which are impossible to solve
analytically even with the program MATHEMATICA. It is more illuminating to
discuss one simple case (i.e. when $c_{3}=0$) in which the constraints
simplify and can be solved explicitly. This corresponds to an
Einstein-Gauss-Bonnet theory. In this case inserting $c_{3}=0$ in Eqs. (\ref%
{def2.5}), (\ref{def3}), (\ref{pol1}) and (\ref{pol2}) one gets
\begin{equation}
c_{0}=\frac{(25c_{1}^{2}+25c_{1}c_{2}\tilde{\eta}+9c_{2}^{2}\tilde{\eta}^{2})%
}{(35c_{2})}
\end{equation}%
\begin{equation}
D_{0}=-\frac{(20c_{1})}{(24c_{2})}  \label{d02}
\end{equation}%
\begin{equation}
\eta =-\frac{(25c_{1})}{(18c_{2})}
\end{equation}%
One can also check that the expression found here for $D_{0}$ in Eq. (\ref%
{d02}) is consistent with the definition (\ref{def3}) and that for this
choice of the coupling constants the theory has two distinct maximally
symmetric eight dimensional vacua with cosmological constants $\Lambda
_{1,2} $
\begin{equation}
\Lambda _{1}=\frac{(-21c_{1}-2\sqrt{14}c_{1})}{(42c_{2})};\;\;\;\Lambda _{2}=%
\frac{(-21c_{1}+2\sqrt{14}c_{1})}{(42c_{2})}
\end{equation}

Indeed, these solutions can be seen as effective vacuum five
dimensional wormholes interpolating between two asymptotic region
where the spatial sections $t=const$ have the same (constant)
curvature (due to the fact that both $g_{tt}$ and $g_{rr}$ are equal
to one in Eq. (\ref{nondewo1})). In the "torsionless" case in five
dimensions, vacuum wormhole solutions have been constructed in the
Chern-Simons case \cite{Tulin}: when such wormholes have the base
manifold of constant curvature, the metric component $g_{tt}$ is not
fixed by the equations of motion (this is a typical sign of the
enhanced gauge symmetry of Chern-Simons theory). However, in the
present non-degenerate case, such degeneracies are completely absent
since all the metric components are fixed by the equations of
motion. In the previous case of the degenerate wormhole (in which
the conditions in Eq. (\ref{degeneracy2}) are fulfilled) the
degeneracies are avoided provided the torsion is non-vanishing in the $%
\Sigma _{2}$ sub-manifold.

\section{How to cross the throat?}

An important issue when dealing with wormholes is their "navigableness".
Indeed one of the most natural questions which arises when dealing with
wormholes is if a timelike or null geodesic can pass through the throat. In
the case that only non-geodesic curves can cross the wormhole means that a
hypothetical astronaut needs an engine to cross the wormhole. It is a known
fact that torsion couples to the spin of a particle. This means that in the
discussion of the "navigableness" of the wormhole one must distinguish
between scalars and spinning particles. This opens the intriguing
possibility that a wormhole with torsion can act as a geometric filter
distinguishing between scalars and spinors and also between the helicities
of particles.

We will therefore begin first to study the "navigableness" for scalar
particles. In the case of nonzero torsion one has two different definitions
of geodesic curve for a scalar which in general do not coincide. The first
possible definition of geodesic is as the curve which extremizes the
particle action
\begin{equation}
I=\int ds^{2}
\end{equation}%
In this definition the torsion does not enter explicitly. The other possible
definition of geodesic is given as an unaccellerated autoparallel curve for
which the connection $\Gamma _{jk}^{k}$ and so the torsion enter directly.
This implies that the two definitions are in general not equivalent. In the
latter case the geodesic equation is given by
\begin{equation}
\ddot{x}^{\mu }+\Gamma _{\nu \lambda }^{\mu }\dot{x}^{\nu }\dot{x}^{\lambda
}=0
\end{equation}%
where the connection symbols can be decomposed as
\begin{equation}
\Gamma _{\nu \lambda }^{\mu }=\hat{\Gamma}_{\nu \lambda }^{\mu }+K_{\nu
\lambda }^{\mu }
\end{equation}%
in which $\hat{\Gamma}_{\nu \lambda }^{\mu }$ are the Christoffel symbols
and $K_{\nu \lambda }^{\mu }$ is the contorsion tensor. In our ansatz for
the torsion in Eq. (\ref{anstor}) the contorsion takes the simple form
\begin{equation}
K_{\nu \lambda }^{\mu }=e_{i}^{\mu }e_{\nu }^{j}K_{j\lambda }^{i}\approx
-e_{i}^{\mu }e_{\nu }^{j}\epsilon _{jk}^{i}e_{\mu }^{k}\approx -\epsilon
_{\nu \lambda }^{\mu }.
\end{equation}%
This implies that the correction to the autoparallel geodesic equation is
identically zero due to the contraction with $\dot{x}^{\nu }\dot{x}^{\lambda
}$. One concludes that for this special form of the contorsion the two
definitions of scalar geodesics coincide.

Naively one should expect that scalar particles should not feel the torsion.
In fact, torsion enters directly the metric through the constant $\delta
_{(1)}$ modifying the size of the throat. Along a geodesic one can normalize
the tangent vector as follows
\begin{equation*}
g_{AB}\left( \partial _{\tau }X^{A}\right) \left( \partial _{\tau
}X^{A}\right) =-k
\end{equation*}%
where $k$ is zero or one for a lightlike and a timelike particle
respectively and we will assume that the tangent vector has no component
along the extra dimensions $a$ (so that $A=1$, $0$, $i$). For the sake of
simplicity, let us consider a geodesic in which only one angular coordinate
(say, $\phi $) is not constant: such a case is enough to show the effects of
torsion on the particles dynamics. The effective equation reads
\begin{equation}
-1=-\dot{t}^{2}+\dot{\rho}^{2}+\left( r_{G}\right) ^{2}\cosh ^{2}(\left(
-D_{0}\right) ^{1/2}\rho )\dot{\phi}^{2}
\end{equation}%
where $r_{G}$ is defined in Eq. (\ref{raga}). There are two Killing vectors $%
\xi _{1}=\partial _{t}$ and $\xi _{2}=\partial _{\phi }$ so there are two
conserved quantities along the geodesic worldline $u_{\mu }$ which are $%
E=\xi _{1}\cdot u$ and $J=\xi _{2}\cdot u$. The radial motion reduces to a
one dimensional problem in an effective potential:
\begin{equation}
\dot{\rho}^{2}+V_{eff}=E^{2};\;\;\;V_{eff}=1+\frac{J^{2}}{\left(
r_{G}\right) ^{2}\cosh ^{2}(\left( -D_{0}\right) ^{1/2}\rho )}.
\label{effpot}
\end{equation}%
It is worth to note that particles with a vanishing angular momentum $J$ do
not feel any potential barrier (the non-trivial term of $V_{eff}$ is purely
centrifugal in nature\footnote{%
On the other hand, wormhole solutions in which $f$ is not constant have, in
general, a non-trivial potential barrier even for purely radial motion: see,
for instance, \cite{Tulin}.}). When $E$ is large enough,%
\begin{equation*}
E^{2}>1+\frac{J^{2}}{\left( r_{G}\right) ^{2}},
\end{equation*}%
timelike geodesic with $J\neq 0$ can cross the wormhole's throat. The
interesting feature which discloses the physical effects of torsion on
scalar particles is that when $\left( r_{G}\right) ^{2}$ becomes larger and
larger (so that the strength of the torsion becomes very large as well) the
centrifugal barrier correspondingly becomes lower and lower: it is then
"easier" for an effective five dimensional geodesic to cross the throat. It
is also interesting to note that the opposite limit is the one in which$\
\left( r_{G}\right) ^{2}$\ is small: in this case, the base manifold becomes
almost teleparellelized\footnote{%
Since the total connection $\left( \omega _{tot}\right) ^{ij}$ almost
vanishes being proportional to $\tilde{\gamma}$ and $\tilde{\gamma}\simeq 0$%
, see Eq. (\ref{toconn}).} and the barrier becomes very high: this "almost"
prevents scalar particles with a $J\neq 0$ from crossing the throat\footnote{%
A smaller and smaller $\left\vert \tilde{\gamma}\right\vert $ would
correspond to make the neck narrower and narrower: the limiting case in
which $\left\vert \tilde{\gamma}\right\vert $\ vanishes (in which case the
amplitude of the neck at the throat vanishes as well) has been called
"space-time horn" in \cite{Tulin} and, strictly speaking, does not
correspond to a wormhole.}.

Let us now see what happens in the case of a spinning particle which has no
components in the extra dimensions. In principle, one should study the
corresponding (classical or quantum) equations of motion to determine the
effects of torsion on the dynamics of spin. However, there is a quite
general qualitative argument which provides one with a clear intuitive
picture of the dynamical effects of a purely axial torsion. It is well known
(see, for a detailed review, \cite{Shapiro}) that, in four dimensions, the
classical equation of motion of a spinning particle with intrinsic angular
momentum $\overrightarrow{J}$\ in a background with torsion represented by
the axial vector $\overrightarrow{S}$ (we will neglect here the effects of
curvature to isolate the torsion contribution) can be written schematically
\begin{equation*}
\partial _{t}\overrightarrow{J}\approx \chi \left( \overrightarrow{J}\times
\overrightarrow{S}\right) ,
\end{equation*}%
(where $\chi $\ is some effective coupling constant) which corresponds to
the following interaction Hamiltonian%
\begin{equation}
H_{st}\approx -\chi \overrightarrow{J}\cdot \overrightarrow{S}.
\label{polar1}
\end{equation}
Thus, on very general grounds, one can say that if the torsion effective
coupling constant $\chi $\ is very large compared to other scales\footnote{%
In the present case one may hope to achieve this condition since the
constant of integration $\delta _{(1)}$ is not constrained by the field
equations.} $\overrightarrow{J}$ "is polarized" by $\overrightarrow{S}$:
states in which $\overrightarrow{J}$\ is parallel to $\overrightarrow{S}$\
have energies much lower than the others states.

Even if the above argument holds in four dimensions, it is known that also
in higher dimensions the coupling of the torsion with spin is similar. Let
us discuss, for instance, the case of the Maxwell Lagrangian in a background
with torsion:%
\begin{eqnarray*}
L &=& \sqrt{-g}d^{D}x\widehat{F}_{\mu \nu }\widehat{F}^{\mu \nu
}=L_{0}+L_{I}, \\
L_{I}. &=& \sqrt{-g}d^{D}x\left( 2F\cdot A\cdot T+A\cdot A\cdot T\cdot
T\right) , \\
\widehat{F}_{\mu \nu } &=&2\nabla _{\left[ \mu \right. }A_{\left. \nu \right]
}=F_{\mu \nu }-T_{\mu \nu }^{\rho }A_{\rho }
\end{eqnarray*}%
where $D$ is the number of spacetime dimensions, $A_{\nu }$ is the gauge
field, $F_{\mu \nu }$ is the torsion free field strength, $L_{0}$ is the
torsion free Maxwell Lagrangian and $L_{I}$ is the spin-torsion interaction
term\footnote{%
Indeed, the above Lagrangian is not satisfactory as a fundamental Lagrangian
for the electromagnetic field since it explicitly breaks gauge invariance
when torsion does not vanish. However, our scope here is only to show how
torsion naturally acts as a "polarizator" on particles with spin. Indeed for
our example one could also have chosen a massive spin one field obtaining
the same conclusions as in the massless case} and obvious contractions have
been understood in $L_{I}$. In the limit in which the torsion is large the
energy of the spin-torsion interaction is%
\begin{equation*}
H_{I}\approx -\sqrt{-g}d^{D}x\left( A\cdot A\cdot T\cdot T\right) .
\end{equation*}%
One can "minimize" the above interaction energy by choosing the polarization
of $A_{\mu }$ which maximizes the (integral of the) product $A\cdot A\cdot
T\cdot T$. \ These kinds of effects appear whenever the Levi-Civita
connection acting on tensor fields is corrected by the torsion.

Thus, the throat can really act as a geometrical filter which distinguishes
scalars and the different polarization states of spinning particles: such
"polarization" effect is maximum at the throat since there the effective
strength $\chi _{eff}$ of the torsion (see Eq. (\ref{dewo2}))%
\begin{equation*}
\chi _{eff}\approx \frac{\delta _{(1)}}{r_{G}\cosh \left( \left(
-D_{0}\right) ^{1/2}\rho \right) }
\end{equation*}%
has a maximum while being very small when $\left\vert \rho \right\vert $\ is
large.

\section{Black Holes}

Let us shortly describe Black hole solutions which correspond to the case $%
f=g$ in which Eqs. (\ref{equ2}) and (\ref{equ3}) becomes identical. The
following simplifications in the curvature two forms occurs
\begin{equation}
\ F=\frac{1}{2}(-f^{2})^{\prime \prime}=\frac{1}{2}Z^{\prime \prime
}\;\;\;;\;\;\;B=\frac{1}{2r}(-f^{2})^{\prime }=\frac{1}{2r}(Z)^{\prime
}\;\;\;;\;\;\;A=-B
\end{equation}%
where we have introduced the function $Z$ defines as
\begin{equation}
Z\equiv \tilde{\gamma}-f^{2}=Dr^{2}
\end{equation}%
Eqs. (\ref{equ2}) and (\ref{equ3}) now read
\begin{gather}
\frac{1}{2r}Z^{\prime }\left[ 20c_{1}+12c_{2}\tilde{\eta}+\frac{Z}{r^{2}}%
\left( 4c_{2}+12c_{3}\tilde{\eta}\right) \right] +140c_{0}  \notag \\
+20c_{1}\tilde{\eta}+\frac{Z}{r^{2}}\left( 20c_{1}+12c_{2}\tilde{\eta}%
\right) =0.  \label{bh0}
\end{gather}

\subsection{The generic case}

In the generic case, one express $\frac{1}{2r}Z^{\prime }$ as rational
function of $\frac{Z}{r^{2}}$:%
\begin{equation}
-\frac{Z^{\prime }}{2r}=\frac{140c_{0}+20c_{1}\tilde{\eta}+\frac{Z}{r^{2}}%
\left( 20c_{1}+12c_{2}\tilde{\eta}\right) }{\left[ 20c_{1}+12c_{2}\tilde{\eta%
}+\frac{Z}{r^{2}}\left( 4c_{2}+12c_{3}\tilde{\eta}\right) \right] }.
\label{bh0.5}
\end{equation}%
The other components of the equations of motion are%
\begin{gather}
\frac{Z^{\prime \prime }}{2}\left[ 20c_{1}+12c_{2}\tilde{\eta}+\frac{Z}{r^{2}%
}\left( 4c_{2}+12c_{3}\tilde{\eta}\right) \right] +\frac{Z^{\prime }}{r}%
\left[ 40c_{1}+24c_{2}\tilde{\eta}\right]  \notag \\
+\left( \frac{Z^{\prime }}{2r}\right) ^{2}\left[ 8c_{2}+24c_{3}\tilde{\eta}%
\right] +\frac{Z}{r^{2}}\left[ 20c_{1}+12c_{2}\tilde{\eta}\right]
+420c_{0}+60c_{1}\tilde{\eta}=\epsilon _{i}=0  \label{bh1}
\end{gather}%
\begin{gather}
\frac{Z^{\prime \prime }}{2}\left[ 20c_{1}+4c_{2}\tilde{\eta}+12\frac{Z}{%
r^{2}}\left( c_{2}+c_{3}\tilde{\eta}\right) \right] +\frac{Z^{\prime }}{r}%
\left[ 60c_{1}+12c_{2}\tilde{\eta}+12\frac{Z}{r^{2}}\left( c_{2}+c_{3}\tilde{%
\eta}\right) \right]  \notag \\
+24\left( \frac{Z^{\prime }}{2r}\right) ^{2}\left[ c_{2}+c_{3}\tilde{\eta}%
\right] +\frac{Z}{r^{2}}\left[ 60c_{1}+12c_{2}\tilde{\eta}\right]
+420c_{0}+20c_{1}\tilde{\eta}=\epsilon _{a}=0  \label{bh2}
\end{gather}%
Taking into account Eq. (\ref{bh0.5}), Eq. (\ref{bh1}) allows to express $%
\frac{Z^{\prime \prime }}{2}$ as a rational function $\frac{Z}{r^{2}}$: let
us call such a function $Y_{(i)}$ to stress that it comes from the $i$-th
component of the equations of motion%
\begin{equation}
\epsilon _{i}=0\Rightarrow \frac{Z^{\prime \prime }}{2}=Y_{(i)}\left( \frac{Z%
}{r^{2}},c_{1},c_{2},c_{3},c_{0}\right) .  \label{bhi}
\end{equation}%
On the other hand, using Eqs. (\ref{bh0.5}) and (\ref{bh2}) one can find a
different expression for $\frac{Z^{\prime \prime }}{2}$ as a rational
function of $\frac{Z}{r^{2}}$: let us call such a function $Y_{(a)}$ to
stress that it comes from the $a$-th component of the equations of motion
\begin{equation}
\epsilon _{a}=0\Rightarrow \frac{Z^{\prime \prime }}{2}=Y_{(a)}\left( \frac{Z%
}{r^{2}},c_{1},c_{2},c_{3},c_{0}\right) .  \label{bha}
\end{equation}%
In the generic case, the two expressions\footnote{%
The explicit expressions for $Y_{(i)}$ and $Y_{(a)}$ are quite long and not
very illuminating but we will not need them in the following.} for $Y_{(i)}$
and $Y_{(a)}$ are different so that one has to impose the consistency
condition%
\begin{equation}
Y_{(i)}=Y_{(a)}.  \label{polz}
\end{equation}%
The above equation can be written as a polynomial of finite degree $N$ in $%
\frac{Z}{r^{2}}$ whose constant coefficients are related to the coupling
constants $c_{i}$ and to $\tilde{\eta}$:
\begin{equation}
Y_{(i)}=Y_{(a)}\Rightarrow \sum^{N}a_{n}\left( \frac{Z}{r^{2}}\right) ^{n}=0.
\label{polz1}
\end{equation}%
This actually implies that $\frac{Z}{r^{2}}$ (being one of the root of a
polynomial with constant coefficients) is constant:
\begin{equation*}
Z=\alpha r^{2}\Rightarrow
\end{equation*}%
\begin{eqnarray}
f^{2} &=&-\alpha r^{2}+\tilde{\gamma},  \label{csbh0} \\
\alpha &<&0,\ \ \ \ \tilde{\gamma}<0,  \notag
\end{eqnarray}%
where $\alpha $ is related to the $c_{i}$ and to $\tilde{\eta}$. This looks
like a Chern-Simons black hole provided both $\alpha $ and $\tilde{\gamma}$\
are negative. However, as it will be shown in a moment, unlike the
Chern-Simons case \cite{CGTW07} torsion appears directly in the metric which
takes the form
\begin{equation}
ds^{2}=-(-\alpha r^{2}+\tilde{\gamma})dt^{2}+\frac{dr^{2}}{(-\alpha r^{2}+%
\tilde{\gamma})}+r^{2}d\Sigma _{1}^{2}+d\Sigma _{2}^{2}.  \label{csbh}
\end{equation}%
The torsion equations in this case give
\begin{equation}
\epsilon _{ij}=0\Rightarrow \alpha (4c_{2}+12c_{3}\tilde{\eta}%
)+20c_{1}+12c_{2}\tilde{\eta}=0
\end{equation}%
\begin{equation}
\epsilon _{ab}=0\Rightarrow \alpha (4c_{2}+12c_{3}\alpha )-24c_{3}\alpha
^{2}+20c_{1}+12c_{2}\alpha =0;
\end{equation}%
such equations put two further constraints on the coefficients $c_{i}$. One
can see that having nonzero torsion in both the $ij$ components as well in
the $ab$ components (that is, $\delta _{(1)}\neq 0$ and $K_{2}\neq 0$ in
Eqs. (\ref{tor1}) and (\ref{tor2})) is not consistent since it would lead to
$c_{3}=c_{2}=0$.

In the case in which torsion is nonzero only in the $ij$ components (that
is, $\delta _{(1)}\neq 0$ in Eq. (\ref{tor1})), one can use the previous
equations to express $c_{0}$, $c_{1}$, $c_{2}$ in function of $\tilde{\eta}%
,\alpha $ and $c_{3}$
\begin{equation}
c_{0}=\frac{3\alpha ^{2}c_{3}\tilde{\eta}(5\alpha ^{2}-5\alpha \tilde{\eta}-4%
\tilde{\eta}^{2})}{35(4\alpha ^{2}-9\alpha \tilde{\eta}+3\tilde{\eta}^{2})}
\end{equation}%
\begin{equation}
c_{1}=\frac{3\alpha ^{2}c_{3}\tilde{\eta}(-5\alpha +7\tilde{\eta})}{%
5(4\alpha ^{2}-9\alpha \tilde{\eta}+3\tilde{\eta}^{2})}
\end{equation}%
\begin{equation}
c_{2}=\frac{3\alpha c_{3}(\alpha -\tilde{\eta})\tilde{\eta}}{4\alpha
^{2}-9\alpha \tilde{\eta}+3\tilde{\eta}^{2}}
\end{equation}%
>From the above expressions for $c_{0}$, $c_{1}$, $c_{2}$, one can see $%
\alpha $ depends on $\tilde{\eta}$: in other words, the lapse function of
the effective five-dimensional black hole depends on the torsion in the
extra-dimensions. It is also worth to stress that $\tilde{\gamma}$ is left
undetermined by the equations of motion. The ratios $\frac{c_{2}^{2}}{%
c_{1}c_{3}}$ and $\frac{c_{1}^{2}}{c_{0}c_{2}}$ are rational functions of $%
\tilde{\eta}$ and $\tilde{\gamma}$ so that these solutions do not belong to
the Born-Infeld class. In the case that one has a nonzero torsion only in
the $ab$ components one obtains
\begin{equation}
c_{0}=\frac{3}{35}\alpha ^{2}c_{3}(5\alpha -4\tilde{\eta})
\end{equation}%
\begin{equation}
c_{1}=-\frac{9\alpha ^{2}c_{3}}{5}
\end{equation}%
\begin{equation}
c_{2}=3\alpha c_{3}
\end{equation}

\subsection{Degenerate black holes}

Black hole solutions in the case in which the degeneracy conditions in Eq. (%
\ref{degeneracy1}) hold are also interesting. In such a case, Eqs. (\ref{bh0}%
), (\ref{bh1}) are identically satisfied and one is only left with Eq. (\ref%
{bh2}):%
\begin{gather}
\frac{Z^{\prime \prime }}{2}\left[ 20c_{1}+4c_{2}\tilde{\eta}+12\frac{Z}{%
r^{2}}\left( c_{2}+c_{3}\tilde{\eta}\right) \right] +\frac{Z^{\prime }}{r}%
\left[ 60c_{1}+12c_{2}\tilde{\eta}+12\frac{Z}{r^{2}}\left( c_{2}+c_{3}\tilde{%
\eta}\right) \right]  \notag \\
+24\left( \frac{Z^{\prime }}{2r}\right) ^{2}\left[ c_{2}+c_{3}\tilde{\eta}%
\right] +\frac{Z}{r^{2}}\left[ 60c_{1}+12c_{2}\tilde{\eta}\right]
+420c_{0}+20c_{1}\tilde{\eta}=0  \label{csbhdeg}
\end{gather}%
Indeed, this is a quite non-trivial non-linear differential equation
which is difficult to solve in general.
Anyway it is easy to show
that effective five dimensional Chern-Simons black holes like in Eq.
(\ref{csbh}) can solve Eq. (\ref{csbhdeg}). If one searches for
Chern-Simons black holes characterized by Eq. (\ref{csbh0}), a
nonzero torsion in the $ab$ components (that is, $K_{2}\neq 0$ in
Eq. (\ref{tor2})) would be inconsistent since it would lead to
$c_{3}=c_{2}=0$. In the case in which torsion is non-vanishing
only in the $ij$ components (that is, $\delta _{(1)}\neq 0$ in Eq. (\ref%
{tor1}) while $K_{2}=0$ so that $\tilde{\eta}=\eta $) one gets
\begin{equation}
\alpha =\frac{-5c_{2}\pm \sqrt{10}\left\vert c_{2}\right\vert }{15c_{3}}
\label{chbhdeg2}
\end{equation}%
\begin{equation}
c_{2}=-3c_{3}\eta .  \label{csbhdeg3}
\end{equation}%
It is interesting to note that, in the case in which $c_{2}$ is positive Eq.
(\ref{chbhdeg2}) implies that $c_{3}$ has to be positive in order to have a
black hole (since $\alpha $ has to be negative) and this would imply that $%
\eta $ (which is the constant curvature of the extra-dimensions) is
negative. Thus, in order to have compact extra-dimensions one has to
compactify the manifold $\Sigma _{2}$ (the standard procedure is to
quotient the hyperbolic three-dimensional constant curvature space
by a freely acting discrete group $\Gamma $).

It remains  the open question of finding more general black hole
solutions in the degenerate case. The most natural ansatz
representing further   black hole solutions would be a compactified
Boulware-Deser black hole \cite{deser} or a compactified
Schwarzschild-(Anti)de-Sitter black hole. However one can check that
those ansatz do not satisfy the above equation of motion. Moreover
even modifying the exponents of the radial coordinate in the lapse
function does not improve the situation. Therefore possible further
black hole solutions will have a structure quite different form from
the one we found. Finding such solutions seems to be a highly
nontrivial but interesting task and will be object of future
investigation.
\section{Generalized Bertotti-Robinson solutions}

For the sake of completeness, here we will shortly describe a class of
generalized Bertotti-Robinson spacetimes. In this case we search for
solutions of the form $\left( A\right) dS_{2}\times S_{3}\times S_{3}$ so
that the metric reads
\begin{equation}
ds^{2}=\frac{l^{2}}{r^{2}}(-dt^{2}+dr^{2})+d\Sigma _{1}^{2}+d\Sigma _{2}^{2}
\end{equation}%
In this ansatz the curvature two forms are
\begin{equation}
R^{0a}=R^{1a}=R^{ia}=R^{0i}=R^{1i}=0
\end{equation}%
\begin{equation}
R^{01}=-\frac{1}{l^{2}}e^{0}e^{1}
\end{equation}

\begin{equation}
R^{ij}=\tilde{\gamma}e^{i}e^{j}\equiv D_{0}e^{i}e^{j}
\end{equation}%
\begin{equation}
R^{ab}=\tilde{\eta}e^{a}e^{b}
\end{equation}%
and the torsion is now%
\begin{equation}
\begin{array}{c}
T^{i}=\delta _{(1)}\epsilon ^{ijk}e_{j}e_{k},\ \ \ \ \ \ T^{a}=K_{2}\epsilon
^{abc}e_{b}e_{c},%
\end{array}%
\end{equation}%
\begin{equation*}
\tilde{\gamma}=\gamma -(\delta _{(1)})^{2},\ \ \ \ \tilde{\eta}=\gamma
-K_{2}^{2},
\end{equation*}%
where, as in the previous sections, $\delta _{(1)}$\ and $K_{2}$\ are
constants. The equations of motion become
\begin{equation*}
\epsilon _{0}=\epsilon _{1}=140c_{0}+20c_{1}\tilde{\gamma}+20c_{1}\tilde{\eta%
}+12c_{2}\tilde{\gamma}\tilde{\eta}=0
\end{equation*}%
\begin{equation*}
\epsilon _{i}=-\frac{1}{l^{2}}\left( 20c_{1}+4c_{2}\tilde{\gamma}+12c_{2}%
\tilde{\eta}+12c_{3}\tilde{\gamma}\tilde{\eta}\right) +420c_{0}+20c_{1}%
\tilde{\gamma}+60c_{1}\tilde{\eta}+12c_{2}\tilde{\gamma}\tilde{\eta}=0
\end{equation*}%
\begin{equation*}
\epsilon _{a}=-\frac{1}{l^{2}}\left( 20c_{1}+4c_{2}\tilde{\eta}+12c_{2}%
\tilde{\gamma}+12c_{3}\tilde{\gamma}\tilde{\eta}\right) +420c_{0}+20c_{1}%
\tilde{\eta}+60c_{1}\tilde{\gamma}+12c_{2}\tilde{\gamma}\tilde{\eta}=0
\end{equation*}%
The torsion equations read
\begin{equation*}
\epsilon _{ij}=\delta _{(1)}\left[ -\frac{1}{l^{2}}\left( 4c_{2}+12c_{3}%
\tilde{\eta}\right) +20c_{1}+12c_{2}\tilde{\eta}\right] =0
\end{equation*}%
\begin{equation*}
\epsilon _{ab}=K_{2}\left[ -\frac{1}{l^{2}}\left( 4c_{2}+12c_{3}\tilde{\gamma%
}\right) +20c_{1}+12c_{2}\tilde{\gamma}\right] =0
\end{equation*}%
When both $\delta _{(1)}$\ and $K_{2}$\ are non-vanishing, the two torsion
equations imply that or $\tilde{\gamma}=\tilde{\eta}$. If one is interested
in the cases in which $\tilde{\gamma}\neq \tilde{\eta}$ then one of the two
torsions is zero (i.e. or $\delta _{(1)}=0$ or $K_{2}=0$)

Let us suppose for the sake of simplicity that only $\delta _{(1)}$\ is
non-vanishing so that $\tilde{\eta}=\eta $. The equations of motion allow to
explicitly specify the couplings $c_{0},c_{1},c_{2},c_{3}$ in terms of $l$, $%
\tilde{\eta}$ and $\tilde{\gamma}$ (so that only three of the four $c_{i}$
are independent\footnote{%
The expressions below have been checked with the software MATEMATICA.}):%
\begin{equation}
c_{3}=-\frac{(\tilde{\gamma}^{2}(1+3\tilde{\gamma}l^{2}+6\eta (-\eta +\tilde{%
\gamma})l^{4}))}{(-2\eta +\tilde{\gamma}+3(-2\eta ^{2}+\tilde{\gamma}%
^{2})l^{2}+12\eta \tilde{\gamma}(-\eta +\tilde{\gamma})l^{4})}  \label{BR1}
\end{equation}%
\begin{equation}
c_{1}=\frac{3\eta \tilde{\gamma}^{2}(\eta +2\tilde{\gamma}+9\eta \tilde{%
\gamma}l^{2})}{5(2\eta -\tilde{\gamma}+3(2\eta ^{2}-\tilde{\gamma}%
^{2})l^{2}+12\eta (\eta -\tilde{\gamma})\tilde{\gamma}l^{4})}  \label{BR2}
\end{equation}%
\begin{equation}
c_{2}=-\frac{3\eta \tilde{\gamma}^{2}(1+(2\eta +\tilde{\gamma})l^{2})}{2\eta
-\tilde{\gamma}+3(2\eta ^{2}-\tilde{\gamma}^{2})l^{2}+12\eta (\eta -\tilde{%
\gamma})\tilde{\gamma}l^{4}}  \label{BR3}
\end{equation}%
\begin{equation}
c_{0}=-\frac{(3\eta \tilde{\gamma}^{2}(\eta ^{2}+2\tilde{\gamma}^{2}+3\eta
\tilde{\gamma}(\eta +2\tilde{\gamma})l^{2}))}{(35(2\eta -\tilde{\gamma}%
+3(2\eta ^{2}-\tilde{\gamma}^{2})l^{2}+12\eta (\eta -\tilde{\gamma})\tilde{%
\gamma}l^{4}))}.  \label{BR4}
\end{equation}%
It is worth to point out that the fractions $\frac{c_{2}^{2}}{c_{1}c_{3}}$
and $\frac{c_{1}^{2}}{c_{0}c_{2}}$ are rational functions of $\eta $ and $%
\tilde{\gamma}$ so that, in general, the above solutions do not belong to
Born-Infeld theory.

In the case in which both torsions are switched on (namely $\delta _{(1)}$\
and $K_{2}$\ non-vanishing) so that $\tilde{\gamma}=\tilde{\eta}$ the above
equations remain valid.

A quite non-trivial characteristic of the present construction is that the
sizes as well as the curvatures of the three factors $\left( A\right) dS_{2}$%
, $S_{3}$, and the second factor $S_{3}$\ can have in principle very
different values as it is clear from Eqs. (\ref{BR1}), (\ref{BR2}), (\ref%
{BR3}) and (\ref{BR4}). Namely, nothing prevents, for instance,
$\eta $ from being of the same order of $\tilde{\gamma}$\ and, at
the same time, the AdS radius $l^{2}$\ from being much larger than
the size of the extra dimensions. The remarkable fact is that this
can be achieved\textit{\ in vacuum}: the only price to pay is that
$c_{3}$ turns out to be much larger
than the other $c_{i}$ in this limit (since only in the expression for $%
c_{3} $ in Eq. (\ref{BR1}) the highest power in $l$ in the numerator is the
same as in the denominator).

\section{Conclusions}

In this paper we constructed exact vacuum solutions with torsion in 8-D
Lovelock theory of the form $M_{5}\times S_{3}$, which can then be seen as
effective five dimensional geometries. The solutions that have been found
are static vacuum effective five-dimensional wormholes, black holes and
generalized Bertotti-Robinson solutions in which the three compact
extra-dimensions play a "spectator" role. All these solutions can carry
nontrivial torsion even in the non-Born-Infeld case. The wormhole
"navigableness" has been discussed and it has been shown that the torsion
has very different effects on scalar or spinning particles. A huge amount of
torsion improves the "navigableness" for scalars while acting as a
"polarizator" on spinning particles.

\section*{Acknowledgements}

The authors would like to thank J. Oliva and G. Giribet for
illuminating discussions and suggestions, the authors would like to
thank R. Troncoso and J. Zanelli for very important comments. This
work was supported by Fondecyt grant 3070055, 3070057 The Centro de
Estudios Cient\'{\i}ficos (CECS) is funded by the Chilean Government
through the Millennium Science Initiative and the Centers of
Excellence Base Financing Program of Conicyt. CECS is also supported
by a group of private companies which at present includes
Antofagasta Minerals, Arauco, Empresas CMPC, Indura, Naviera
Ultragas and Telef\'{o}nica del Sur. CIN is funded by Conicyt and
the Gobierno Regional de Los R\'{\i}os. F.C. gratefully acknowledges
the Agenzia Spaziale Italiana for partial support.

\end{document}